# *Multiple-relaxation-time lattice Boltzmann model for simulating double-diffusive convection in fluid-saturated porous media*[1]

Qing Liu, Ya-Ling He

*Key Laboratory of Thermo-Fluid Science and Engineering of Ministry of Education, School of Energy and Power Engineering, Xi'an Jiaotong University, Xi'an, Shaanxi, 710049, China*

**Abstract**

Double-diffusive convection in porous media is a common phenomenon in nature, and has received considerable attention in a wide variety of engineering applications. In this paper, a multiple-relaxation-time (MRT) lattice Boltzmann (LB) model is developed for simulating double-diffusive convection in porous media at the representative elementary volume scale. The MRT-LB model is constructed in the framework of the triple-distribution-function approach: the velocity field, the temperature and concentration fields are solved separately by three different MRT-LB equations. The present model has two distinctive features. First, the equilibrium moments of the temperature and concentration distributions have been modified, which makes the effective thermal diffusivity and heat capacity ratio as well as the effective mass diffusivity and porosity decoupled. This feature is very useful in practical applications. Second, source terms have been added into the MRT-LB equations of the temperature and concentration fields so as to recover the macroscopic temperature and concentration equations. Numerical tests demonstrate that the present model can serve as an accurate and efficient numerical method for simulating double-diffusive convection in porous media.

---

*media*; *heat and mass transfer*.

## 1. Introduction

Double-diffusive convection (also called the thermosolutal convection) in fluid-saturated porous media has attracted a great deal of attention because it is a common phenomenon in nature, and it is also frequently encountered in a wide variety of engineering applications, such as geophysical systems, drying processes, geophysical systems, the migration of moisture contained in fibrous insulation, the transport of contaminants in groundwater, the underground disposal of nuclear wastes and so on [1-7]. Over the last several decades, double-diffusive convection in porous media has been studied numerically by many researchers. Conventional numerical methods, such as the finite-element method (FEM) [1,2], the finite-volume method (FVM) [3,4], and the finite-difference method (FDM) [5], have been employed to study double-diffusive convection problems in porous media. Comprehensive reviews of the subject have been given by Ingham and Pop [6], and Nield and Bejan [7]. The above mentioned studies [1-5] were carried out using conventional numerical methods based on the discretization of the macroscopic continuum equations. In order to get a thorough understanding of the underlying mechanisms, more fundamental approaches should be developed for studying double-diffusive convection in porous media systems.

The lattice Boltzmann (LB) method, as a mesoscopic numerical technique originated from the lattice-gas automata (LGA) method [8], has achieved great success in simulating complex fluid flows and modeling complex physics in fluids [9-16]. Unlike the conventional numerical methods, which are based on the discretization of macroscopic continuum equations, the LB method simulates fluid flows by tracking the evolution of the single-particle distribution function based on the mesoscopic kinetic equation. Owing to its kinetic background, the LB method exhibits some attractive advantages over the convectional numerical methods [17]: (i) non-linearity (collision process) is completely local and non-locality (streaming process) is completely linear, whereas the transport term $\mathbf{u}\cdot\nabla\mathbf{u}$ in the Navier-Stokes (N-S) equations is non-linear and non-local at a time; (ii) the pressure of the LB method is simply calculated by an equation of state, while in convectional numerical methods it is usually necessary and costly to solve a Poisson equation for the pressure field of the incompressible N-S equations; (iii) complex boundary conditions can be easily formulated in terms of elementary mechanical rules; (iv) nearly ideal for parallel computing with very low communication/computation ratio.

The LB method has also been successfully applied to study double-diffusive convection problems. In 2002, Guo et al. [18] proposed an LB model for double-diffusive natural convection system. Guo et al.'s model was constructed in the framework of the triple-distribution-function (TDF) approach, and it can predict the flow and heat/mass transfer characteristics correctly. Later, Lu and Zhan [19] investigated the three-dimensional thermosolutal convection flow in a cubic cavity. The limitations of the LB method in simulating double-diffusive convection were discussed. Verhaeghe et al. [20] have presented an LB method to study double-diffusive natural convection phenomena induced by temperature and concentration gradients in a multicomponent fluid in two and three dimensions. In their method, the temperature field was solved by a finite-difference scheme. In the literature [21], Yu et al. proposed an LB model to study double-diffusive natural convection in a horizontal shallow cavity. In their study, the Dufour and Soret effects on the heat and mass transfer processes have been considered by modifying the equilibrium distribution functions of the temperature and concentration, respectively. Mohamad et al. [22] have studied double-diffusive natural convection in an open-ended cavity for a wide range of characteristic parameters. The physics of double-diffusive natural convection due to temperature and concentration buoyancy forces were explored and discussed with opposite temperature and concentration gradients. In addition, Nazari et al. [23] have simulated double-diffusive convection in a square cavity in the presence of a hot square obstacle. The results showed that the Nusselt and Sherwood numbers increase with the increase of the Rayleigh number and aspect ratio.

Moreover, Zhao et al. [24] investigated the double-diffusive convection in a square cavity filled with a porous medium at the pore scale. In Zhao et al.'s study, the Dufour and Soret effects on the heat and mass transfer processes were considered by adding additional source terms into the LB equations of the temperature and concentration fields, respectively. Later, Xu et al. [25] investigated the double-diffusive convection around a heated cylinder in a square cavity filled with porous medium at the representative elementary volume (REV) scale. Xu et al.'s study was based on the Brinkman-extended Darcy model. Recently, Chen et al. [26] have developed an LB model for simulating double-diffusive convection in fluid-saturated porous media at the REV scale. Chen et al.'s model was developed based on the generalized non-Darcy model, and by introducing a reference porosity into the equilibrium concentration distribution, it can work well not only for uniform porous media but also for non-uniform porous media.

Up to now, although some progress has been made in studying double-diffusive convection in fluid-saturated porous media at the REV scale, some drawbacks of the LB method are apparent. For instance, the effective thermal diffusivity and heat capacity ratio as well as the effective mass diffusivity and porosity is coupled. The artificial couplings may do harm to the accuracy and numerical stability of the LB method. Therefore, the objective of this work is to propose an LB model for simulating double-diffusive convection in porous media, in which the above-mentioned drawback will be overcome. Considering that the multiple-relaxation-time (MRT) collision model [27,28] is superior to its Bhatnagar-Gross-Krook (BGK) counterpart [29] in terms of accuracy and numerical stability, the MRT collision model is employed in this work.

The rest of this paper is organized as follows. The macroscopic governing equations for double-diffusive convection in porous media at the REV scale are briefly described in Section 2. Section 3 presents the MRT-LB model for double-diffusive convection in porous media in detail. In Section 4, numerical simulations are carried out to demonstrate the accuracy and reliability of the MRT-LB model. Finally, some conclusions are given in Section 5.

**2. Macroscopic governing equations**

Based on the generalized non-Darcy model [30-32], the macroscopic governing equations for double-diffusive convection in porous media at the REV scale can be written as follows [1-3,7]:

$$\nabla \cdot \mathbf{u} = 0, \tag{1}$$

$$\frac{\partial \mathbf{u}}{\partial t} + (\mathbf{u} \cdot \nabla)\left(\frac{\mathbf{u}}{\phi}\right) = -\frac{1}{\rho_0}\nabla(\phi p) + \upsilon_e \nabla^2 \mathbf{u} + \mathbf{F}, \tag{2}$$

$$\sigma \frac{\partial T}{\partial t} + \mathbf{u} \cdot \nabla T = \nabla \cdot (\alpha_e \nabla T), \tag{3}$$

$$\phi \frac{\partial C}{\partial t} + \mathbf{u} \cdot \nabla C = \nabla \cdot (D_e \nabla C), \tag{4}$$

where $\mathbf{u}$, $p$, $T$ and $C$ are the volume-averaged velocity, pressure, temperature and concentration, respectively; $\rho_0$ is the reference density, $\phi$ is the porosity, $\upsilon_e$ is the effective viscosity, $\alpha_e$ is the effective thermal diffusivity, and $D_e$ is the effective mass diffusivity; $\sigma$ is the heat capacity ratio given by $\sigma = \phi + (1-\phi)(\rho_s c_{ps})/(\rho_f c_{pf})$, in which $\rho_f$ ($\rho_s$) and $c_{pf}$ ($c_{ps}$) are the density and specific heat of the fluid (solid matrix), respectively; $\mathbf{F} = (F_x, F_y)$ is the total body force induced by the porous matrix and other external forces, which can be expressed as [33,34]

$$\mathbf{F} = -\frac{\phi \upsilon}{K}\mathbf{u} - \frac{\phi F_\phi}{\sqrt{K}}|\mathbf{u}|\mathbf{u} + \phi \mathbf{G}, \tag{5}$$

where $\upsilon$ is the viscosity of the fluid ($\upsilon$ is not necessarily the same as $\upsilon_e$); $K$ and $F_\phi$ are the permeability and inertial coefficient (Forchheimer coefficient) of the porous medium, respectively; $|\mathbf{u}| = \sqrt{u_x^2 + u_y^2}$, in which $u_x$ and $u_y$ are the components of $\mathbf{u}$ in the $x$- and $y$-directions, respectively. The buoyancy force $\mathbf{G}$ is given by

$$\mathbf{G} = g\left[\beta_T(T - T_0) + \beta_C(C - C_0)\right]\mathbf{j}, \tag{6}$$

where $T_0$ and $C_0$ are the reference temperature and concentration, respectively; $\beta_T$ and $\beta_C$ are the thermal and concentration expansion coefficients, respectively; $g$ is the gravitational acceleration, and $\mathbf{j}$ is the unit vector in the $y$-direction.

The inertial coefficient $F_\phi$ and permeability $K$ depend on the geometry of the porous media. For flow over a packed bed of particles, according to Ergun's experimental investigations [35], $F_\phi$ and $K$ can be expressed as [36]

$$F_\phi = \frac{1.75}{\sqrt{150\phi^3}}, \quad K = \frac{\phi^3 d_p^2}{150(1-\phi)^2}, \tag{7}$$

where $d_p$ is the solid particle diameter (or mean pore diameter).

The system governed by Eqs. (1)-(4) is characterized by several dimensionless characteristic parameters: the Prandtl number $Pr = \upsilon/\alpha_e$, the thermal Rayleigh number $Ra = g\beta_T \Delta T L^3/(\upsilon \alpha_e)$ (thermal Grashof number $Gr = g\beta_T \Delta T L^3/\upsilon^2$), the solutal Rayleigh number $Ra_C = g\beta_C \Delta C L^3/(\upsilon D_e)$ (solutal Grashof number $Gr_C = g\beta_C \Delta C L^3/\upsilon^2$), the Darcy number $Da = K/L^2$, the viscosity ratio $J = \upsilon_e/\upsilon$, the Reynolds number $Re = LU/\upsilon$, the Lewis number $Le = \alpha_e/D_e$, the Schmidt number $Sc = \upsilon/D_e$, and the buoyancy ratio $N = \beta_C \Delta C/(\beta_T \Delta T)$, where $L$ is the characteristic length, $U$ is the characteristic velocity, $\Delta T$ is the temperature difference (characteristic temperature), and $\Delta C$ is the concentration difference (characteristic concentration).

3. **MRT-LB model for double-diffusive convection in porous media**

The MRT method [27,28] is an important extension of the relaxation LB method developed by

Higuera et al. [37]. In the MRT method, the collision process of the LB equation is executed in moment space, while the streaming process of the LB equation is carried out in velocity space. By using the MRT collision model, the relaxation times of the hydrodynamic and non-hydrodynamic moments can be separated. The MRT method has become more and more popular since a detailed theoretical analysis was made by Lallemand and Luo [28] in 2000. In what follows, an MRT-LB model for double-diffusive convection in porous media at the REV scale will be presented. The model is constructed in the framework of the TDF approach: the velocity field, the temperature and concentration fields are solved separately by three different MRT-LB equations.

*3.1. MRT-LB equation for the velocity field*

For the velocity field, the two-dimensional nine-velocity (D2Q9) lattice is employed. The nine discrete velocities $\{\mathbf{e}_i\}$ of the D2Q9 lattice are given by [29]

$$\mathbf{e}_i = \begin{cases} (0,0), & i = 0, \\ \left(\cos\left[(i-1)\pi/2\right], \sin\left[(i-1)\pi/2\right]\right)c, & i = 1 \sim 4, \\ \left(\cos\left[(2i-9)\pi/4\right], \sin\left[(2i-9)\pi/4\right]\right)\sqrt{2}c, & i = 5 \sim 8, \end{cases} \quad (8)$$

where $c = \delta_x/\delta_t$ is the lattice speed with $\delta_t$ and $\delta_x$ being the discrete time step and lattice spacing, respectively. The lattice speed $c$ is set to be 1 ($\delta_x = \delta_t$) in this work.

According to Refs. [38-41], the MRT-LB equation with an explicit treatment of the forcing term can be expressed as

$$f_i\left(\mathbf{x}+\mathbf{e}_i\delta_t, t+\delta_t\right) = f_i(\mathbf{x},t) - \tilde{\Lambda}_{ij}\left(f_j - f_j^{eq}\right)\Big|_{(\mathbf{x},t)} + \delta_t\left(\tilde{S}_i - 0.5\tilde{\Lambda}_{ij}\tilde{S}_j\right), \quad (9)$$

where $f_i$ is the (volume-averaged) density distribution function, $f_i^{eq}$ is the equilibrium distribution function, $\tilde{\Lambda} = \mathbf{M}^{-1}\Lambda\mathbf{M}$ is the collision matrix, and $\tilde{S}_i$ is the forcing term, in which $\mathbf{M}$ is the transformation matrix, and $\Lambda = \text{diag}(s_\rho, s_e, s_\varepsilon, s_j, s_q, s_j, s_q, s_v, s_v)$ is the relaxation matrix. The transformation matrix $\mathbf{M}$ linearly maps the discrete distribution functions represented by $\mathbf{f} \in \mathbb{V} = \mathbb{R}^9$ to their (velocity) moments represented by $\mathbf{m} \in \mathbb{M} = \mathbb{R}^9$: $\mathbf{m} = \mathbf{M}\mathbf{f}$ and $\mathbf{f} = \mathbf{M}^{-1}\mathbf{m}$, where $\mathbf{f}$ and $\mathbf{m}$ denote nine-dimensional column vectors, e.g., $\mathbf{f} = (f_0, f_1, \ldots, f_8)^{\mathrm{T}}$.

Through the transformation matrix $\mathbf{M}$, the collision process of the MRT-LB equation (9) can be executed in moment space, i.e.,

$$\mathbf{m}^*(\mathbf{x}, t) = \mathbf{m}(\mathbf{x}, t) - \Lambda(\mathbf{m} - \mathbf{m}^{eq})\big|_{(\mathbf{x}, t)} + \delta_t \left(\mathbf{I} - \frac{\Lambda}{2}\right)\mathbf{S}, \tag{10}$$

where $\mathbf{S} = \mathbf{M}\tilde{\mathbf{S}}$. The streaming process is still carried out in velocity space

$$f_i(\mathbf{x} + \mathbf{e}_i \delta_t, t + \delta_t) = f_i^*(\mathbf{x}, t), \tag{11}$$

where $\mathbf{f}^* = \mathbf{M}^{-1}\mathbf{m}^*$.

For the D2Q9 model, the nine moments corresponding to the discrete velocities are

$$\mathbf{m} = \left(\rho, e, \varepsilon, j_x - \frac{\delta_t}{2}\rho F_x, q_x, j_y - \frac{\delta_t}{2}\rho F_y, q_y, p_{xx}, p_{xy}\right)^{\mathrm{T}}, \tag{12}$$

where $\mathbf{J} = (j_x, j_y) = \rho(u_x, u_y)$ is the flow momentum. With the ordering of the moments specified as the above, the transformation matrix $\mathbf{M}$ can be explicitly given by [28]

$$\mathbf{M} = \begin{bmatrix} 1 & 1 & 1 & 1 & 1 & 1 & 1 & 1 & 1 \\ -4 & -1 & -1 & -1 & -1 & 2 & 2 & 2 & 2 \\ 4 & -2 & -2 & -2 & -2 & 1 & 1 & 1 & 1 \\ 0 & 1 & 0 & -1 & 0 & 1 & -1 & -1 & 1 \\ 0 & -2 & 0 & 2 & 0 & 1 & -1 & -1 & 1 \\ 0 & 0 & 1 & 0 & -1 & 1 & 1 & -1 & -1 \\ 0 & 0 & -2 & 0 & 2 & 1 & 1 & -1 & -1 \\ 0 & 1 & -1 & 1 & -1 & 0 & 0 & 0 & 0 \\ 0 & 0 & 0 & 0 & 0 & 1 & -1 & 1 & -1 \end{bmatrix}. \tag{13}$$

The equilibrium moment vector $\mathbf{m}^{eq}$ corresponding to $\mathbf{m}$ is defined as [41]

$$\mathbf{m}^{eq} = \left[\rho, -2\rho + \frac{3\rho_0 |\mathbf{u}|^2}{\phi}, \alpha_1 \rho + \frac{\alpha_2 \rho_0 |\mathbf{u}|^2}{\phi}, \rho_0 u_x, -\rho_0 u_x, \rho_0 u_y, -\rho_0 u_y, \frac{\rho_0 (u_x^2 - u_y^2)}{\phi}, \frac{\rho_0 u_x u_y}{\phi}\right]^{\mathrm{T}}, \tag{14}$$

where $\alpha_1$ and $\alpha_2$ are free parameters of the model. The components of the forcing term $\mathbf{S}$ in the moment space are given as follows [41]

$$S_0 = 0, \quad S_1 = \frac{6\rho_0 \mathbf{u} \cdot \mathbf{F}}{\phi}, \quad S_2 = -\frac{6\rho_0 \mathbf{u} \cdot \mathbf{F}}{\phi}, \quad S_3 = \rho_0 F_x, \quad S_4 = -\rho_0 F_x,$$

$$S_5 = \rho_0 F_y, \quad S_6 = -\rho_0 F_y, \quad S_7 = \frac{2\rho_0(u_x F_x - u_y F_y)}{\phi}, \quad S_8 = \frac{\rho_0(u_x F_y + u_y F_x)}{\phi}. \tag{15}$$

Note that the incompressibility approximation, i.e., $\rho = \rho_0 + \delta\rho \approx \rho_0$ ($\delta\rho$ is the density fluctuation) and $\mathbf{J} = (j_x, j_y) \approx \rho_0 \mathbf{u}$, have been employed in Eqs. (14) and (15).

The macroscopic density $\rho$ and velocity $\mathbf{u}$ are given by

$$\rho = \sum_{i=0}^{8} f_i, \tag{16}$$

$$\rho_0 \mathbf{u} = \sum_{i=0}^{8} \mathbf{e}_i f_i + \frac{\delta_t}{2} \rho_0 \mathbf{F}. \tag{17}$$

The macroscopic pressure $p$ is defined by $p = \rho c_s^2 / \phi$, and the effective viscosity $\upsilon_e$ is given by $\upsilon_e = c_s^2 (s_\upsilon^{-1} - 0.5) \delta_t$, where $c_s = c/\sqrt{3}$ is the sound speed of the D2Q9 model. It should be noted that the total body force $\mathbf{F}$ also contains the velocity $\mathbf{u}$, so Eq. (17) is a nonlinear equation for $\mathbf{u}$. According to Ref. [34], the macroscopic velocity can be calculated explicitly by

$$\mathbf{u} = \frac{\mathbf{v}}{l_0 + \sqrt{l_0^2 + l_1 |\mathbf{v}|}}, \tag{18}$$

where

$$\rho_0 \mathbf{v} = \sum_{i=0}^{8} \mathbf{e}_i f_i + \frac{\delta_t}{2} \phi \rho_0 \mathbf{G}, \quad l_0 = \frac{1}{2}\left(1 + \phi \frac{\delta_t}{2} \frac{\upsilon}{K}\right), \quad l_1 = \phi \frac{\delta_t}{2} \frac{F_\phi}{\sqrt{K}}. \tag{19}$$

The equilibrium distribution function $f_i^{eq}$ in velocity space is given by ($\alpha_1 = 1$, $\alpha_2 = -3$)

$$f_i^{eq} = w_i \left\{ \rho + \rho_0 \left[ \frac{\mathbf{e}_i \cdot \mathbf{u}}{c_s^2} + \frac{(\mathbf{e}_i \cdot \mathbf{u})^2}{2\phi c_s^4} - \frac{|\mathbf{u}|^2}{2\phi c_s^2} \right] \right\}, \tag{20}$$

where $w_0 = 4/9$, $w_{1\sim 4} = 1/9$, and $w_{5\sim 8} = 1/36$.

### 3.2. MRT-LB equations for the temperature and concentration fields

For double-diffusive convection in porous media, the temperature and concentration fields governed by Eqs. (3) and (4) are solved separately by two different MRT-LB equations based on the two-dimensional five-velocity (D2Q5) lattice. The five discrete velocities $\{\mathbf{e}_i | i = 0 \sim 4\}$ of the D2Q5

lattice are given in Eq. (8). The MRT-LB equations for the temperature and concentration fields are given by

$$\mathbf{g}(\mathbf{x}+\mathbf{e}\delta_t,t+\delta_t) = \mathbf{g}(\mathbf{x},t) - \mathbf{N}^{-1}\Theta(\mathbf{n}_g - \mathbf{n}_g^{eq})\big|_{(\mathbf{x},t)} + \delta_t \mathbf{N}^{-1}\left(\mathbf{I} - \frac{\Theta}{2}\right)\mathbf{S}^T, \tag{21}$$

$$\mathbf{h}(\mathbf{x}+\mathbf{e}\delta_t,t+\delta_t) = \mathbf{h}(\mathbf{x},t) - \mathbf{N}^{-1}\mathbf{Q}(\mathbf{n}_h - \mathbf{n}_h^{eq})\big|_{(\mathbf{x},t)} + \delta_t \mathbf{N}^{-1}\left(\mathbf{I} - \frac{\mathbf{Q}}{2}\right)\mathbf{S}^C, \tag{22}$$

respectively, where the bold-face symbols ($\mathbf{g}$, $\mathbf{h}$, $\mathbf{n}_g$, $\mathbf{n}_h$, $\mathbf{S}^T$, and $\mathbf{S}^C$) denote five-dimensional column vectors of moments, e.g., $\mathbf{g} = (g_0,\ldots,g_4)^\top$; $g_i(\mathbf{x},t)$ and $h_i(\mathbf{x},t)$ are the temperature and concentration distribution functions, respectively; $\mathbf{N}$ is the transformation matrix, $\Theta = \mathrm{diag}(\zeta_0,\zeta_\alpha,\zeta_\alpha,\zeta_e,\zeta_\varepsilon)$ and $\mathbf{Q} = \mathrm{diag}(\eta_0,\eta_D,\eta_D,\eta_e,\eta_\varepsilon)$ are relaxation matrices.

The transformation matrix $\mathbf{N}$ linearly maps the discrete distribution functions represented by $\mathbf{g}$ and $\mathbf{h}$ to their moments represented by $\mathbf{n}_g$ and $\mathbf{n}_h$: $\mathbf{n}_g = \mathbf{N}\mathbf{g}$ and $\mathbf{n}_h = \mathbf{N}\mathbf{h}$. Through the transformation matrix $\mathbf{N}$, the collision processes of the MRT-LB equations (21) and (22) can be executed in moment space $\mathbb{M} = \mathbb{R}^5$, while the streaming processes are still carried out in velocity space $\mathbb{V} = \mathbb{R}^5$. For the D2Q5 model, the transformation matrix $\mathbf{N}$ can be chosen as [42,43]

$$\mathbf{N} = \begin{bmatrix} 1 & 1 & 1 & 1 & 1 \\ 0 & 1 & 0 & -1 & 0 \\ 0 & 0 & 1 & 0 & -1 \\ -4 & 1 & 1 & 1 & 1 \\ 0 & 1 & -1 & 1 & -1 \end{bmatrix}. \tag{23}$$

Note that $\mathbf{N}$ is an orthogonal transformation matrix. The non-orthogonal transformation matrix [44,45] can also be employed in the present study. The equilibrium moment vectors $\mathbf{n}_g^{eq}$ and $\mathbf{n}_h^{eq}$ are chosen as follows

$$\mathbf{n}_g^{eq} = (\sigma T, u_x T, u_y T, -4\sigma T + 5\varpi T, 0)^\top, \tag{24}$$

$$\mathbf{n}_h^{eq} = (\phi C, u_x C, u_y C, -4\phi C + 5\varpi C, 0)^\top, \tag{25}$$

where $\varpi \in (0,1)$ is a model parameter.

The macroscopic temperature and concentration can be obtained via

$$\sigma T = \sum_{i=0}^{4} g_i, \quad \phi C = \sum_{i=0}^{4} h_i. \tag{26}$$

In order to recover the macroscopic equations (3) and (4), according to Refs. [46-48], the source terms can be chosen as $\mathbf{S}^T = \mathbf{N}\left(\tilde{S}_0^T, \ldots, \tilde{S}_4^T\right)^{\mathrm{T}}$ and $\mathbf{S}^C = \mathbf{N}\left(\tilde{S}_0^C, \ldots, \tilde{S}_4^C\right)^{\mathrm{T}}$ with

$$\tilde{S}_i^T = \tilde{w}_i \frac{\mathbf{e}_i}{c_{sT}^2} \cdot \frac{\partial (T\mathbf{u})}{\partial t}, \quad \tilde{S}_i^C = \tilde{w}_i \frac{\mathbf{e}_i}{c_{sT}^2} \cdot \frac{\partial (C\mathbf{u})}{\partial t}, \tag{27}$$

where $\{\tilde{w}_i\}$ are the weight coefficients: $\tilde{w}_0 = 1 - \varpi$ and $\tilde{w}_{1\sim 4} = \varpi/4$, and $c_{sT} = \sqrt{\varpi/2}$ is the sound speed of the D2Q5 model ($c_{sT}^2 = \sum_i \tilde{w}_i \mathbf{e}_i^2 / 2$ [17]).

Through the Chapman-Enskog analysis of the MRT-LB equations (21) and (22), the macroscopic temperature and concentration equations (3) and (4) can be correctly recovered. Without the source terms, there must exist unwanted terms ($\nabla \cdot \left[\delta_t \left(\zeta_\alpha^{-1} - 0.5\right) \epsilon \partial_{t_1} (T\mathbf{u})\right]$ and $\nabla \cdot \left[\delta_t \left(\eta_D^{-1} - 0.5\right) \epsilon \partial_{t_1} (C\mathbf{u})\right]$) in the macroscopic equations. In simulations, the explicit difference scheme can be used to compute the time derivative terms (e.g., $\partial_t (T\mathbf{u}) = \left[(T\mathbf{u})\big|_{(\mathbf{x},t)} - (T\mathbf{u})\big|_{(\mathbf{x},t-\delta_t)}\right]/\delta_t$), such a treatment does not affect the inherent merits of the LB method. The equilibria $g_i^{eq}$ and $h_i^{eq}$ in velocity space are given by

$$g_i^{eq} = \begin{cases} \sigma T - (1 - \tilde{w}_0) T, & i = 0, \\ \tilde{w}_i T \left(1 + \frac{\mathbf{e}_i \cdot \mathbf{u}}{c_{sT}^2}\right), & i = 1 \sim 4, \end{cases} \quad h_i^{eq} = \begin{cases} \phi C - (1 - \tilde{w}_0) C, & i = 0, \\ \tilde{w}_i C \left(1 + \frac{\mathbf{e}_i \cdot \mathbf{u}}{c_{sT}^2}\right), & i = 1 \sim 4, \end{cases} \tag{28}$$

respectively.

The effective thermal diffusivity $\alpha_e$ and effective mass diffusivity $D_e$ are defined as

$$\alpha_e = c_{sT}^2 \left(\frac{1}{\zeta_\alpha} - \frac{1}{2}\right) \delta_t, \quad D_e = c_{sT}^2 \left(\frac{1}{\eta_D} - \frac{1}{2}\right) \delta_t, \tag{29}$$

respectively.

In the thermal LB models (the temperature field is governed by Eq. (3)) [41,45,49], the effective thermal diffusivity $\alpha_e$ depends on the heat capacity ratio $\sigma$. Usually, following the line of Guo and Zhao's model [49], the equilibrium moments of the temperature-based and concentration-based MRT-LB models can be defined as

$$\mathbf{n}_g^{eq} = \left[\sigma T, u_x T, u_y T, (-4+5\varpi)\sigma T, 0\right]^\top, \tag{30}$$

$$\mathbf{n}_h^{eq} = \left[\phi C, u_x C, u_y C, (-4+5\varpi)\phi C, 0\right]^\top. \tag{31}$$

The equilibria $g_i^{eq}$ and $h_i^{eq}$ in velocity space are given by $g_i^{eq} = \tilde{w}_i T\left(\sigma + \mathbf{e}_i \cdot \mathbf{u}/c_{sT}^2\right)$ and $h_i^{eq} = \tilde{w}_i C\left(\phi + \mathbf{e}_i \cdot \mathbf{u}/c_{sT}^2\right)$, respectively. The effective diffusivities $\alpha_e$ and $D_e$ are given by

$$\alpha_e = \sigma c_{sT}^2 \left(\frac{1}{\zeta_\alpha} - \frac{1}{2}\right)\delta_t, \quad D_e = \phi c_{sT}^2 \left(\frac{1}{\eta_D} - \frac{1}{2}\right)\delta_t. \tag{32}$$

Obviously, for the temperature and concentration fields governed by Eqs. (3) and (4), the heat capacity ratio $\sigma$ and porosity $\phi$ only influence the solutions of the transient state. Theoretically, for double-diffusive convection in porous media at the REV scale governed by Eqs. (1)-(4), the steady-state solutions are independent of $\sigma$ and $\phi$. In numerical studies of steady-state double-diffusive convection in porous media using convectional numerical methods, the transient terms ($\sigma \partial_t T$ and $\phi \partial_t C$) can be directly dropped (e.g., Ref. [4]). The LB method is not a method of choice for steady-state computations [50], and therefore the heat capacity ratio $\sigma$ and porosity $\phi$ should be incorporated into the equilibrium moments of the temperature and concentration distributions. However, the dependence of effective thermal diffusivity on heat capacity ratio as well as the dependence of effective mass diffusivity on porosity (see Eq. (32)) is not necessary. As shown in Eqs. (24), (25) and (29), by modifying the equilibrium moments appropriately, the artificial coupling between the effective thermal diffusivity and heat capacity ratio as well as the artificial coupling between the effective mass diffusivity and porosity has been avoided.

### *3.3. Boundary condition treatments*

In what follows, the boundary condition treatments are briefly introduced. For velocity and temperature/concentration boundary conditions, the non-equilibrium extrapolation scheme [51] is adopted. For a boundary node $\mathbf{x}_b$ where $\mathbf{u}(\mathbf{x}_b, t)$ is known, but $\rho(\mathbf{x}_b, t)$ is unknown, the density

distribution function $f_i(\mathbf{x}_b, t)$ at the boundary node $\mathbf{x}_b$ is approximated by

$$f_i(\mathbf{x}_b, t) = \tilde{f}_i^{eq}(\mathbf{x}_b, t) + \left[ f_i(\mathbf{x}_f, t) - f_i^{eq}(\mathbf{x}_f, t) \right], \tag{33}$$

where $\tilde{f}_i^{eq}(\mathbf{x}_b, t) = f_i^{eq}(\rho(\mathbf{x}_f), \mathbf{u}(\mathbf{x}_b), t)$, in which $\mathbf{x}_f$ is the nearest neighbor fluid node of $\mathbf{x}_b$ along the link $\mathbf{e}_i$ (i.e., $\mathbf{x}_f = \mathbf{x}_b + \mathbf{e}_i \delta_t$).

If the temperature $T(\mathbf{x}_b, t)$ and concentration $C(\mathbf{x}_b, t)$ at the boundary node $\mathbf{x}_b$ are known, the temperature and concentration distributions $g_i(\mathbf{x}_b, t)$ and $h_i(\mathbf{x}_b, t)$ can be determined by

$$g_i(\mathbf{x}_b, t) = g_i^{eq}(\mathbf{x}_b, t) + \left[ g_i(\mathbf{x}_f, t) - g_i^{eq}(\mathbf{x}_f, t) \right], \tag{34}$$

$$h_i(\mathbf{x}_b, t) = h_i^{eq}(\mathbf{x}_b, t) + \left[ h_i(\mathbf{x}_f, t) - h_i^{eq}(\mathbf{x}_f, t) \right], \tag{35}$$

respectively. When the temperature gradient (or heat flux) and concentration gradient (or mass flux) at the boundary node $\mathbf{x}_b$ are known, the temperature and concentration at the boundary node $\mathbf{x}_b$ can be approximated by the following extrapolation scheme

$$T(\mathbf{x}_b, t) = \frac{4T(\mathbf{x}_f, t) - T(\mathbf{x}_{ff}, t) - 2\Delta\mathbf{x} \cdot \nabla T(\mathbf{x}_b, t)}{3}, \tag{36}$$

$$C(\mathbf{x}_b, t) = \frac{4C(\mathbf{x}_f, t) - C(\mathbf{x}_{ff}, t) - 2\Delta\mathbf{x} \cdot \nabla C(\mathbf{x}_b, t)}{3}, \tag{37}$$

where $\Delta\mathbf{x} = \mathbf{x}_f - \mathbf{x}_b = \mathbf{x}_f - \mathbf{x}_{ff}$ ($\mathbf{x}_{ff}$ is the nearest fluid node of $\mathbf{x}_f$ along the link $\mathbf{e}_i$).

## 4. Numerical simulations

In this section, numerical simulations of several 2D benchmark problems are performed to demonstrate the accuracy and reliability of the MRT-LB model proposed in Section 3. In simulations, we set $\rho_0 = 1$, $J = 1$, $\sigma = 1$, $\delta_x = \delta_y = \delta_t = 1$ ($c_s = 1/\sqrt{3}$), $\varpi = 2/5$ ($c_{sT} = 1/\sqrt{5}$), and $\alpha_e/\alpha = 1$ ($\alpha$ is the thermal diffusivity of the fluid). The free relaxation rates are chosen as follows: $s_0 = s_3 = s_5 = 1$, $s_1 = s_2 = 1.1$, $s_4 = s_6 = 1.2$, $\zeta_0 = 1$, $\zeta_3 = \zeta_4 = 1.5$, $\eta_0 = 1$, and $\eta_3 = \eta_4 = 1.5$.

### *4.1. Double-diffusive convection in a rectangular enclosure without porous media*

When $\phi \to 1$ and $Da$ tends to infinity, the present MRT-LB model reduces to an MRT-LB

model for double-diffusive convection in the absence of a porous medium. In this subsection, the present model is used to simulate the double-diffusive convection flow in a rectangular enclosure without a porous medium (see Fig. 1). The height and width of the rectangular enclosure are $H$ and $L$ (aspect ratio $A = H/L$), respectively. The temperatures ($T_h$ and $T_l$) and concentrations ($C_h$ and $C_l$) are uniformly imposed on the vertical walls ($T_h > T_l$, $C_h > C_l$), while the top and bottom walls are thermally insulated and impermeable to mass transfer. The temperature difference, concentration difference, reference temperature, and reference concentration are $\Delta T = T_h - T_l$, $\Delta C = C_h - C_l$, $T_0 = (T_h + T_l)/2$, and $C_0 = (C_h + C_l)/2$, respectively. The average Nusselt number $\overline{Nu}$ and average Sherwood number $\overline{Sh}$ at the left (or right) vertical wall are defined by

$$\overline{Nu} = \int_0^H Nu(y)\,dy/H, \quad \overline{Sh} = \int_0^H Sh(y)\,dy/H, \tag{38}$$

where $Nu(y) = -L(\partial T/\partial x)_{wall}/\Delta T$ and $Sh(y) = -L(\partial C/\partial x)_{wall}/\Delta C$ are the local Nusselt number and local Sherwood number, respectively.

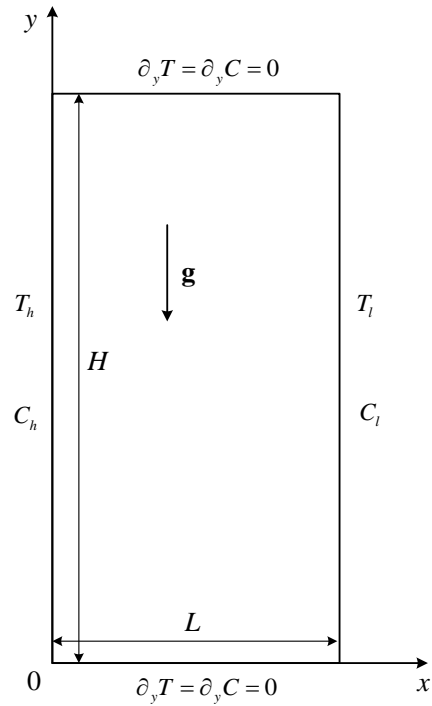

Fig. 1. Physical model of double-diffusive convection in a rectangular enclosure.

In simulations, the main characteristic parameters are set as: $Ra = 10^5$, $Pr = 1$, $Le = 2$, $Da = 10^8$, $\phi = 1$, $A = 2$, and $N = -0.8$. A grid size of $N_x \times N_y = 100 \times 200$ is employed. The relaxation rates $s_\upsilon$, $\zeta_\alpha$ and $\eta_D$ can be fully determined in terms of $Pr$, $Ra$, $Le$ and $Ma$, where $Ma = U/c_s$ is the Mach number and is set to be 0.1 in simulations ($U = \sqrt{\beta g \Delta T L}$ is the characteristic velocity). The relaxation rate $\tau_\upsilon$ is determined by $s_\upsilon^{-1} = 0.5 + MaJL\sqrt{Pr}/\left(c_s \delta_t \sqrt{Ra}\right)$ (in lattice unit, $L = N_x$). The isoconcentrations, streamlines, and isotherms predicted by the present model are shown in Fig. 2. These plots are in good agreement with those reported in Refs. [52,53]. In Fig. 3, the concentration and temperature profiles along the horizontal line crossing the center of the rectangular enclosure are presented. The FEM results [52] are also included for comparison. It can be seen that the present results are in good agreement with the results reported in the literature. To quantify the results, the maximum horizontal velocity component $|u|_{max}$, the maximum vertical velocity component $|v|_{max}$, the average Nusselt number $\overline{Nu}$ and Sherwood number $\overline{Sh}$ at the left vertical wall are measured and listed in Table 1. Obviously, the present numerical results agree well with the results reported in previous studies [53,54].

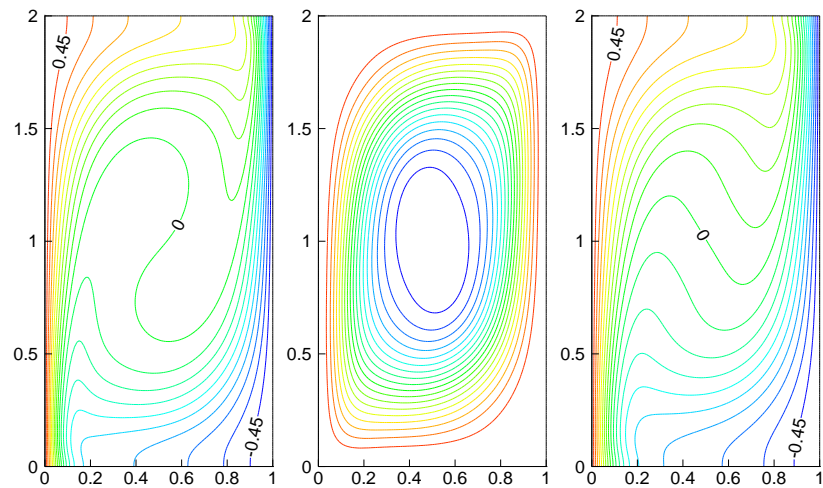

Fig. 2. Isoconcentrations (left), streamlines (middle), and isotherms (right) of double-diffusive convection in a rectangular enclosure ($Da = 10^8$, $\phi = 1$, $Ra = 10^5$, $Pr = 1$, $Le = 2$, $A = 2$,

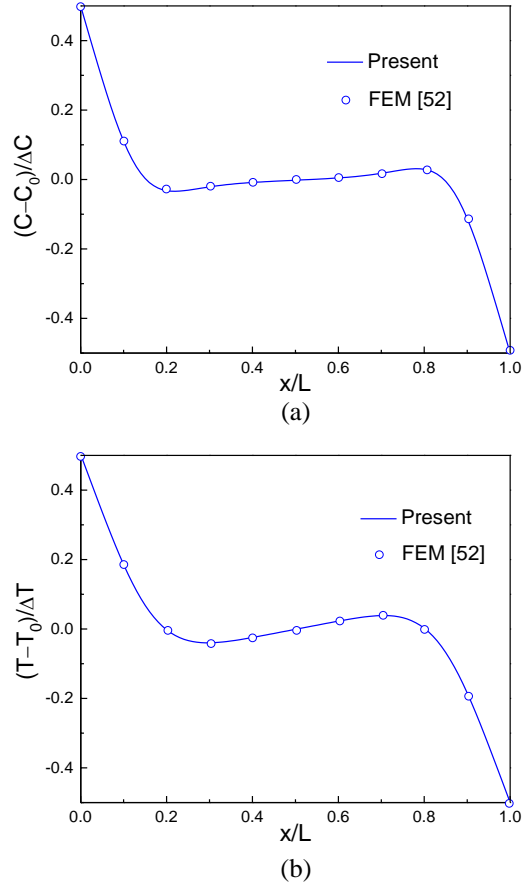

Fig. 3. Concentration (a) and temperature (b) profiles along the horizontal line crossing the center of the rectangular enclosure ($Da = 10^8$, $\phi = 1$, $Ra = 10^5$, $Pr = 1$, $Le = 2$, $A = 2$, and $N = -0.8$).

Symbols: FEM [52]; solid lines: the MRT-LB results.

Table 1 Comparison of the present numerical results with those in Refs. [53,54] ($Da = 10^8$, $\phi = 1$, $Ra = 10^5$, $Pr = 1$, $Le = 2$, $A = 2$, and $N = -0.8$).

|  | $|u_x|_{max}$ | $|u_y|_{max}$ | $\overline{Nu}$ | $\overline{Sh}$ |
| --- | --- | --- | --- | --- |
| FDM [53] | 38.4931 | 86.3858 | 3.4323 | 4.4238 |
| FVM [54] | - | - | 3.45 | 4.48 |
| Present | 39.275 | 86.375 | 3.402 | 4.398 |

*4.2. Double-diffusive natural convection in a porous cavity with a freely convective wall*

Double-diffusive natural convection flow in a porous cavity with freely convective wall conditions has been studied by FEM [2] based on the generalized non-Darcy model. The physical model of this problem is shown in Fig. 4. The left vertical wall of the cavity is maintained at temperature $T_h$ and concentration $C_h$, while the right vertical wall is subjected to convective heat and mass transfer conditions. The top and bottom walls are adiabatic and impermeable to mass transfer. $T_{inf}$ is the ambient temperature, $C_{inf}$ is the ambient concentration, $h_T$ is the convective heat transfer coefficient, and $h_C$ is the convective mass transfer coefficient. The temperature difference, concentration difference, reference temperature, and reference concentration are $\Delta T = T_h - T_{inf}$, $\Delta C = C_h - C_{inf}$, $T_0 = (T_h + T_{inf})/2$, and $C_0 = (C_h + C_{inf})/2$, respectively. The convective heat and mass transfer conditions on the right vertical wall are described by the following relations

$$\frac{\partial \theta}{\partial X} = Bi_T \theta, \quad \frac{\partial \Phi}{\partial X} = Bi_C \Phi, \tag{39}$$

where $\theta = (T - T_{inf})/\Delta T$, $\Phi = (C - C_{inf})/\Delta C$, $X = x/L$, $Y = y/L$, $Bi_T = h_T L/\alpha_e$ is the thermal Biot number, and $Bi_C = h_C L/D_e$ is the solutal Biot number.

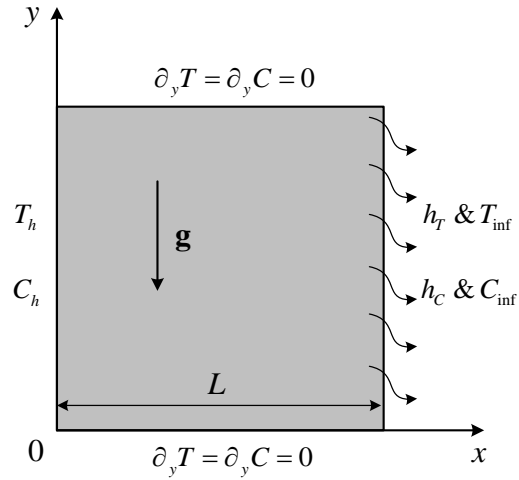

Fig. 4. Physical model of double-diffusive natural convection flow in a porous cavity with freely convective wall.

In this test, we set $Ra = 10^4$, $Da = 10^{-2}$, $Bi_T = 1$, $\phi = 0.6$, $Pr = 1$, $Le = 1$, and $N = 1$. A grid size of $100 \times 100$ is employed in the computations. The isoconcentrations, streamlines, and isotherms for different solutal Biot numbers are shown in Fig. 5. It can be seen that for a small value of solutal

Biot number ($Bi_C = 1$), the isoconcentrations and isotherms do not differ from each other because the strength of the thermal buoyancy force is equal to the solutal buoyancy force ($N = 1$). As $Bi_C$ increases to $200$, the convective effect is significantly enhanced and the flow is much stronger than that at $Bi_C = 1$. These observed phenomena are in good agreement with Ref. [2]. To quantify the comparison, the maximum absolute values of the stream function and the minimum values of the temperature and concentration are measured and listed in Table 2. It can be observed that the numerical results predicted by the present model are in excellent agreement with the numerical results in previous studies [2,3].

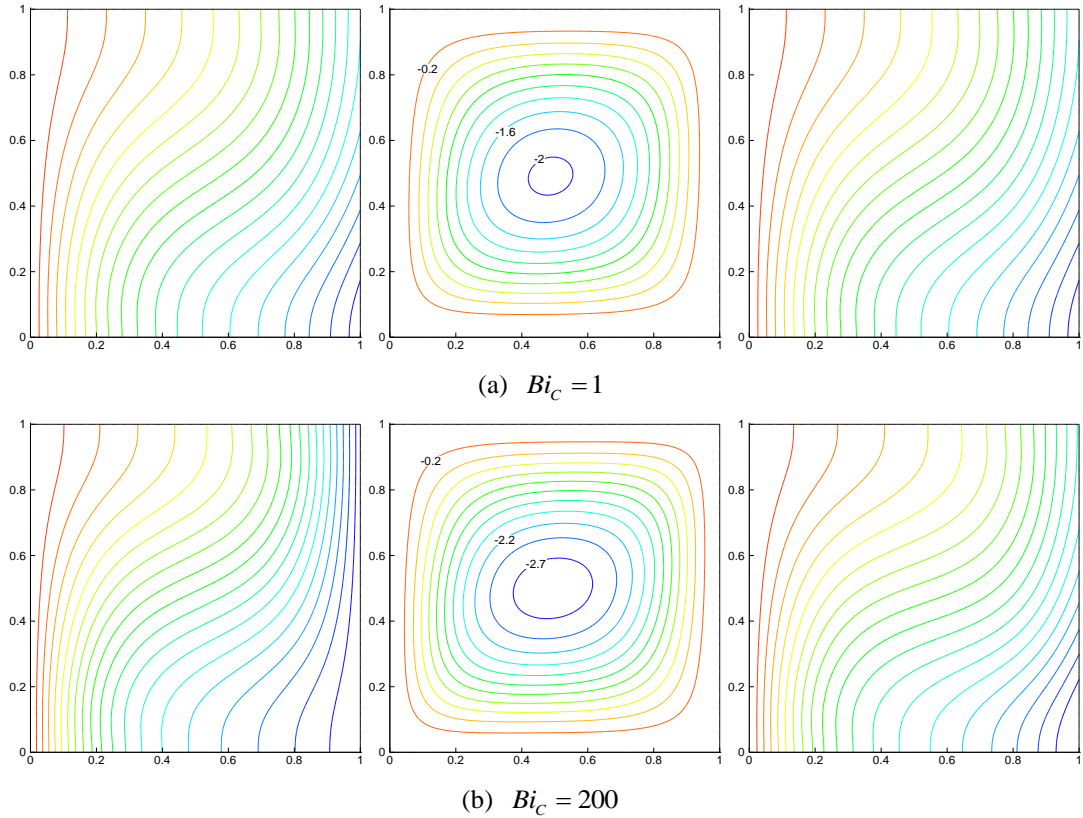

(a) $Bi_C = 1$

(b) $Bi_C = 200$

Fig. 5. Isoconcentrations (left), streamlines (middle), and isotherms (right) for different solutal Biot numbers.

Table 2 Comparison of the present results of $|\psi|_{max}$, $\theta_{min}$, and $\Phi_{min}$ with those reported in Refs. [2,3] for $Bi_C = 1$ and $Bi_C = 200$.

|  | $Bi_C = 1$ | | | $Bi_C = 200$ | | |
| --- | --- | --- | --- | --- | --- | --- |
|  | FEM [2] | FVM [3] | Present | FEM [2] | FVM [3] | Present |
| $|\psi|_{max}$ | 2.042 | 2.041 | 2.03466 | 2.854 | 2.846 | 2.83609 |
| $\theta_{min}$ | 0.504 | 0.5089 | 0.50858 | 0.538 | 0.5412 | 0.54096 |
| $\Phi_{min}$ | 0.504 | 0.5089 | 0.50858 | 0.002 | 0.0021 | 0.00260 |

In Table 3, the minimum values of the temperature and concentration, the average Nusselt number $\overline{Nu}$ and Sherwood number $\overline{Sh}$ at the left vertical wall are listed for $Bi_C = 1$ and $Le = 1$. As discussed in Section 3.2, with the given parameters, the temperature and concentration fields should be identical at the steady state. As shown in Table 3, the minimum values of the temperature and concentration as well as the average Nusselt and Sherwood numbers predicted by the present model are identical, while differences exist between the corresponding results predicted by model-A. Here, "model-A" denotes the equlibria and effective diffusivities are given by Eqs. (30)-(32). Indeed, the artificial couplings (see Eq. (32)) affect the accuracy of the LB method and should be avoided in simulations.

Table 3 Comparison of the present results of $\theta_{min}$, $\Phi_{min}$, $\overline{Nu}$, and $\overline{Sh}$ with those predicted by model-A for $Bi_C = 1$.

|  | $\theta_{min}$ | $\Phi_{min}$ | $\overline{Nu}$ | $\overline{Sh}$ |
| --- | --- | --- | --- | --- |
| FEM [2] | 0.504 | 0.504 | 0.600 | 0.600 |
| FVM [3] | 0.5089 | 0.5089 | - | - |
| Present | 0.508577 | 0.508577 | 0.591041 | 0.591041 |
| Model-A | 0.508618 | 0.508409 | 0.591101 | 0.590609 |

*4.3. Double-diffusive mixed convection in a lid-driven porous cavity*

In this subsection, the present MRT-LB model is employed to simulate the double-diffusive mixed convection flow in a lid-driven cavity filled with porous media, which has been studied by Khanafer and Vafai [3] using FVM based on the generalized non-Darcy model. The physical model of this problem is shown in Fig. 6. The domain under consideration is a 2D square porous cavity where uniform temperatures and concentrations are imposed at the horizontal walls ($T_h$ and $C_h$ at the top wall, and $T_l$ and $C_l$ at the bottom wall), and the vertical walls are adiabatic and impermeable to mass transfer. The top wall moves from left to right with a constant velocity $U_0$, while the other three walls are fixed. The temperature difference, concentration difference, reference temperature, and reference concentration are $\Delta T = T_h - T_l$, $\Delta C = C_h - C_l$, $T_0 = (T_h + T_l)/2$, and $C_0 = (C_h + C_l)/2$, respectively. The Reynolds number is defined by $Re = LU_0/\upsilon$, and the relaxation rate $s_\upsilon$ can be determined by $s_\upsilon^{-1} = 0.5 + JLU_0/(c_s^2 \delta_t Re)$.

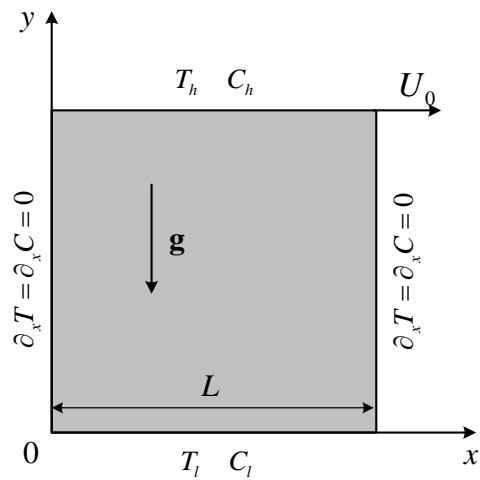

Fig. 6. Physical model of double-diffusive mixed convection flow in a lid-driven porous cavity.

In simulations, unless otherwise specified, we set $Gr = 10^2$, $\phi = 0.95$, $A = 1$, $Sc = 1$, and $U_0 = 0.1$. A grid size of $N_x \times N_y = 128 \times 128$ is employed for all the calculations. The effect of $Da$ on the isoconcentrations, streamlines, and isotherms for $Re = 500$, $Pr = 0.1$, $Le = 10$, and $N = 1$ is shown in Fig. 7. It can be seen that the temperature field is conduction dominated except the domain near the top wall, where the mechanical influence of the lid-driven wall is significant. As $Da$ decreases to $10^{-3}$, the isotherms are almost parallel to the horizontal wall, and the streamlines crowd near the top wall. Moreover, Fig. 7 clearly shows that $Da$ has significant influence on the isoconcentrations, which is more pronounced at smaller values of $Da$.

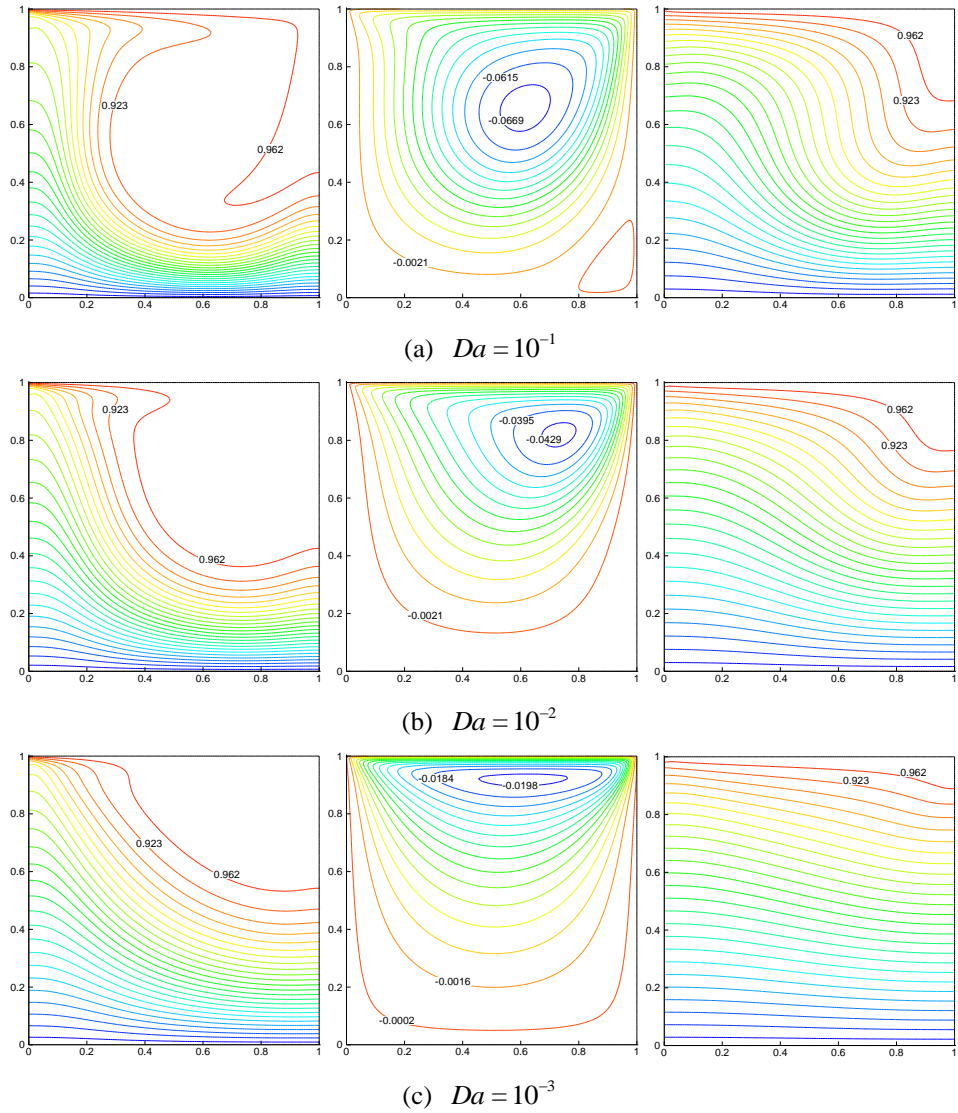

(a) $Da = 10^{-1}$

(b) $Da = 10^{-2}$

(c) $Da = 10^{-3}$

Fig. 7. Effect of $Da$ on the isoconcentrations (left), streamlines (middle), and isotherms (right) for $Re = 500$, $Pr = 0.1$, $Le = 10$ and $N = 1$.

Fig. 8 illustrates the effect of $Re$ on the isoconcentrations, streamlines, and isotherms for $Da = 0.1$, $Pr = 1$, $Le = 1$ and $N = 1$. It is clear that the buoyancy influence is overwhelmed by the mechanical influence of the lid-driven wall as the Reynolds number $Re$ increases. For small value of Reynolds number ($Re = 50$), the isoconcentrations and isotherms are almost parallel to the horizontal wall near the bottom surface. As $Re$ increases to $500$, the mechanical influence is significantly enhanced and the temperature and concentration gradients are small in most of the interior domain of the cavity because the fluid is well mixed inside the cavity. Moreover, the thermal (or concentration) boundary layer is thinner near the bottom wall due to the higher temperature (or concentration) gradient as $Re$ increases.

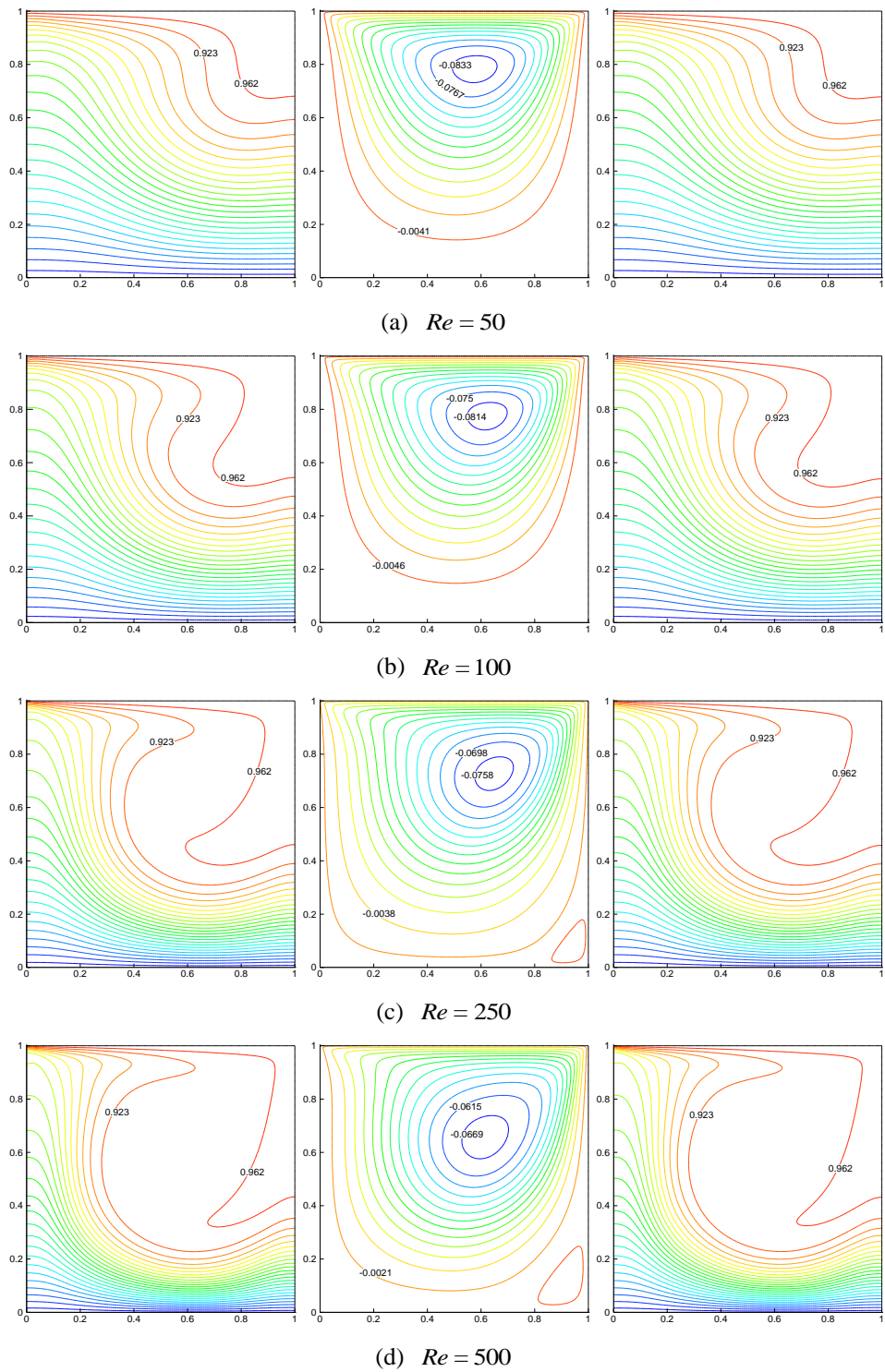

(a) $Re = 50$

(b) $Re = 100$

(c) $Re = 250$

(d) $Re = 500$

Fig. 8. Effect of $Re$ on the isoconcentrations (left), streamlines (middle), and isotherms (right) for $Da = 0.1$, $Pr = 1$, $Le = 1$ and $N = 1$.

Figs. 9 and 10 illustrate the effect of $N$ on the isoconcentrations, streamlines, and isotherms for $Re = 100$, $Da = 0.1$, $Pr = 1$, and $Le = 1$. Fig. 9 clearly shows that as $N$ increases to a high positive value ($N = 500$), the mechanical influence of the lid-driven wall is overwhelmed by the combined influence of the buoyancy forces, and the fluid flow is almost stagnant inside the cavity except in the region close to the lid-driven wall. As shown in Fig. 10, for $N < 0$, the influence of the solutal buoyancy force is acting in the same direction as the mechanical influence of the lid-driven top wall, which causes the vortex (with clockwise rotation) inside the cavity to rotate at higher velocity. Due to the combined effect of the solutal buoyancy force and the lid-driven top wall, the fluid inside the cavity is well mixed, and the thermal and solutal activities inside the cavity are significantly enhanced as $N$ increases to a high negative value ($N = -1500$). To sum up, all the observations from Figs. 7-10 agree well with those reported in Ref. [3].

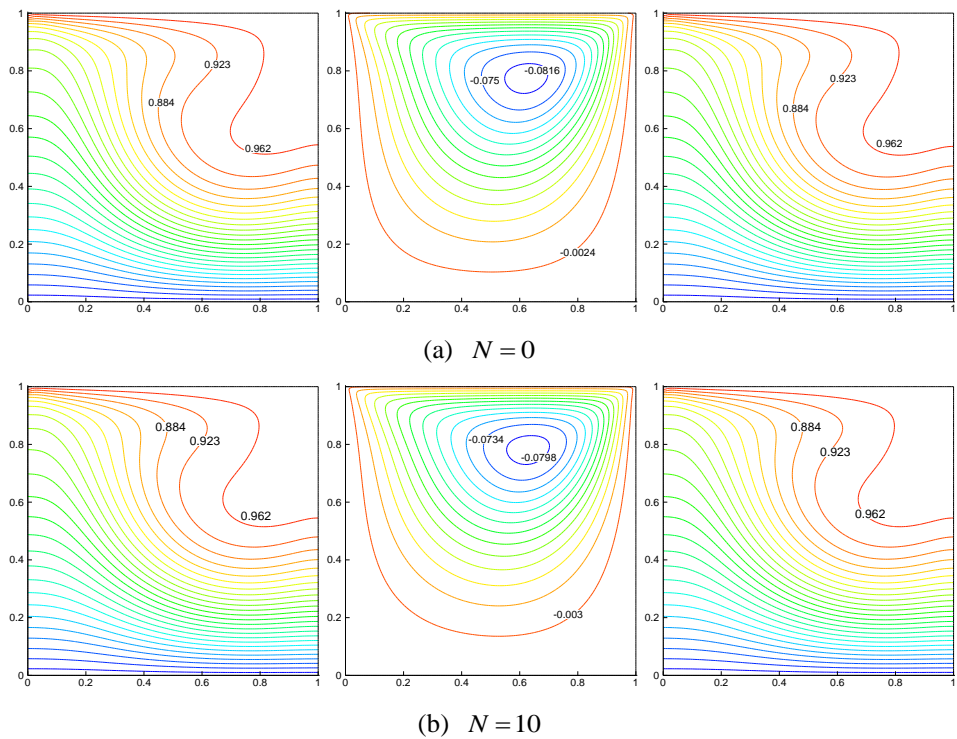

(a) $N = 0$

(b) $N = 10$

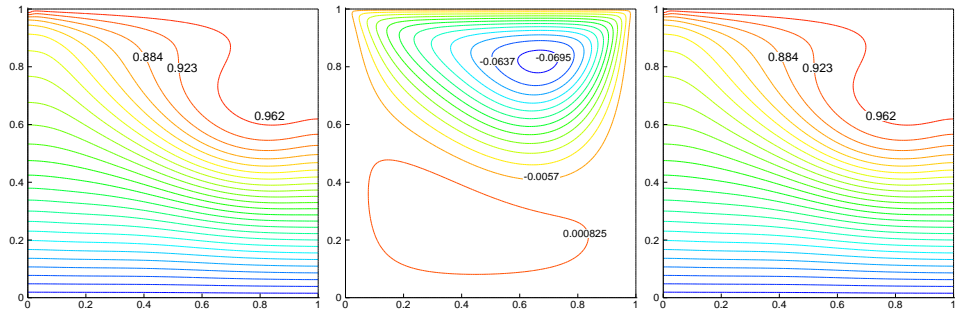

(c) $N = 100$

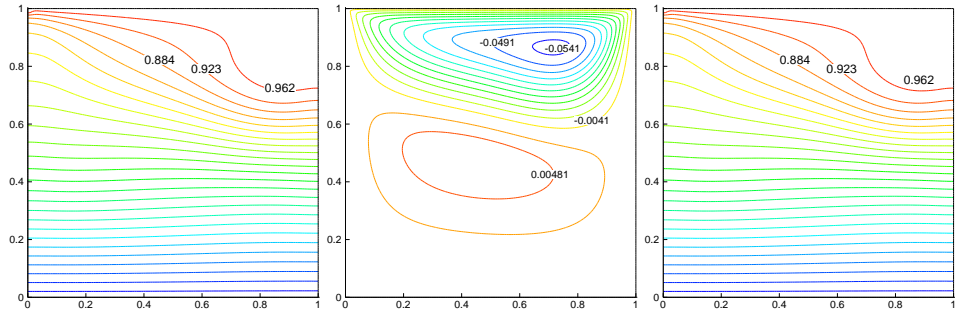

(d) $N = 500$

Fig. 9. Effect of positive (buoyancy ratio values) $N$ on the isoconcentrations (left), streamlines (middle), and isotherms (right) for $Re = 100$, $Da = 0.1$, $Pr = 1$, and $Le = 1$.

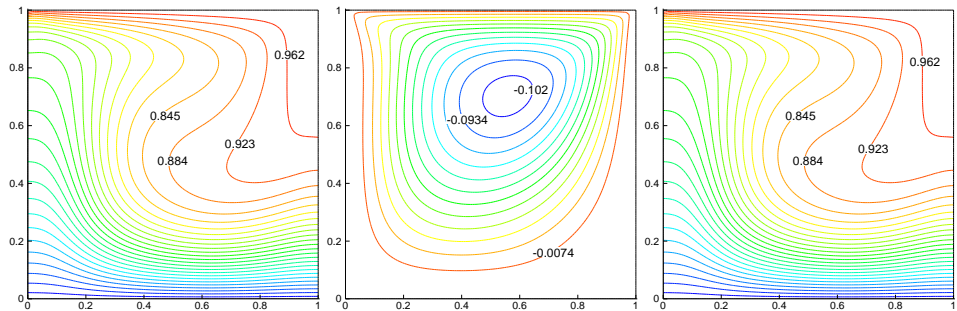

(a) $N = -100$

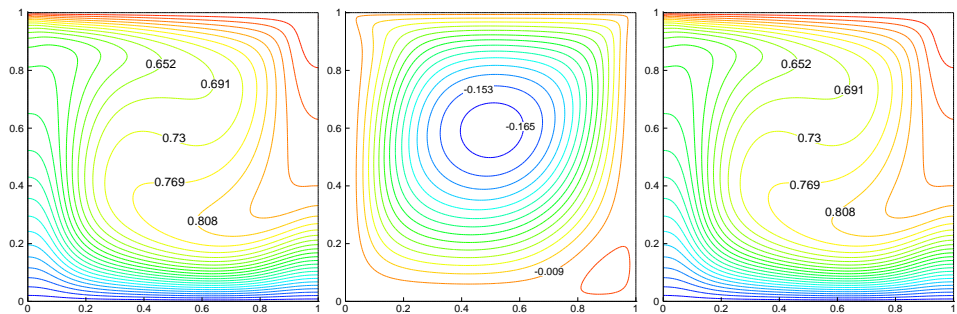

(b) $N = -500$

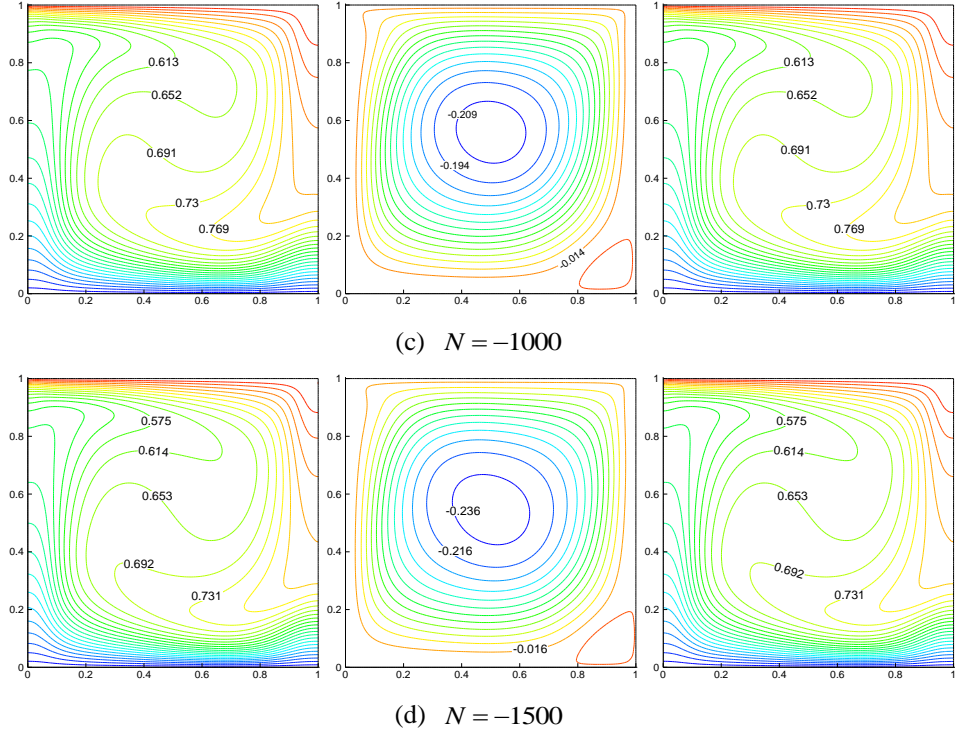

(c) $N = -1000$

(d) $N = -1500$

Fig. 10. Effect of negative (buoyancy ratio values) $N$ on the isoconcentrations (left), streamlines (middle), and isotherms (right) for $Re = 100$, $Da = 0.1$, $Pr = 1$, and $Le = 1$.

To quantify the results, the velocity and temperature profiles along the horizontal line crossing the center of the cavity are compared with the results of Ref. [3] in Fig. 11. It can be observed that the solutions of the present MRT-LB model are in good agreement with the numerical results in Ref. [3]. Moreover, the average Nusselt number $\overline{Nu}$ and average Sherwood number $\overline{Sh}$ at the bottom wall of the cavity for different values of $N$ are measured and listed in Table 4. The numerical results in Ref. [3] are also included in Table 4 for comparison. Obviously, the present numerical results are in excellent agreement with those reported in the literature.

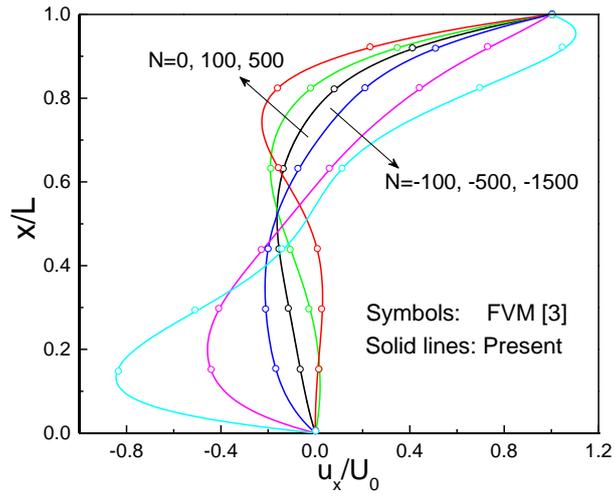

(a)

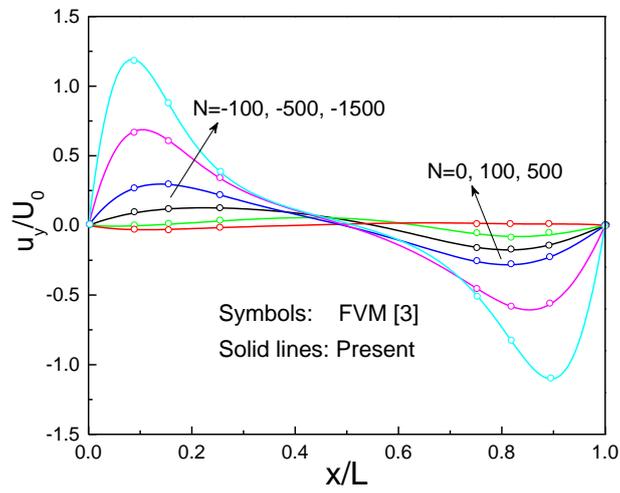

(b)

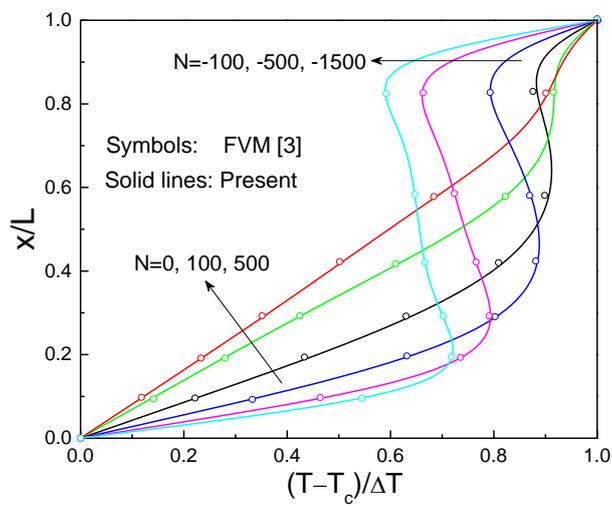

(c)

Fig. 11. Effect of $N$ on the horizontal velocity profile (a), vertical velocity profile (b), and temperature profile (c) along the horizontal line crossing the center of the cavity for $Re=100$, $Da=0.1$, $Pr=1$, and $Le=1$.

Table 4 Comparison of the present results of the average Nusselt and Sherwood numbers with those in Ref. [3] ($Gr=10^2$, $\phi=0.95$, $Re=100$, $Da=0.1$, $Pr=1$, and $Le=1$).

|  |  | \multicolumn{9}{c}{$N$} |
|---|---|---|---|---|---|---|---|---|---|---|
|  |  | -1500 | -1000 | -750 | -500 | -250 | -100 | 0 | 100 | 500 |
| $\overline{Sh}$ | FVM[3] | 5.00 | 4.62 | 4.34 | 4.00 | 3.44 | 2.89 | 2.11 | 1.48 | 1.24 |
|  | Present | 5.0103 | 4.6281 | 4.3611 | 3.9971 | 3.4424 | 2.8914 | 2.1009 | 1.4778 | 1.2208 |
| $\overline{Nu}$ | FVM[3] | 5.00 | 4.62 | 4.34 | 4.00 | 3.44 | 2.89 | 2.11 | 1.48 | 1.24 |
|  | Present | 5.0103 | 4.6281 | 4.3611 | 3.9971 | 3.4424 | 2.8914 | 2.1009 | 1.4778 | 1.2208 |

In Table 5, the average Nusselt number $\overline{Nu}$ and Sherwood number $\overline{Sh}$ at the bottom wall of the cavity are listed ($Gr=10^2$, $\phi=0.6$, $Re=100$, $Da=0.1$, $Pr=1$, and $Le=1$). As shown in Table 5, the average Nusselt and Sherwood numbers predicted by the present model are identical, while the results predicted by model-A are not identical. Theoretically, the average Nusselt and Sherwood numbers should be identical at the steady state. In summary, the present model is more suitable for simulating Double-diffusive convection in porous media at the REV scale.

Table 5 Comparison of the present results of $\overline{Nu}$ and $\overline{Sh}$ with those predicted by model-A ($Gr=10^2$, $\phi=0.6$, $Re=100$, $Da=0.1$, $Pr=1$, and $Le=1$).

| $N$ |  | $\overline{Nu}$ | $\overline{Sh}$ |
|---|---|---|---|
| -1500 | Present | 4.53420 | 4.53420 |
|  | Model-A | 4.54694 | 4.54905 |
| -500 | Present | 3.61929 | 3.61929 |
|  | Model-A | 3.62571 | 3.62634 |

## 5. Conclusions

In this paper, an MRT-LB model for simulating double-diffusive convection in porous media at the REV scale is developed in the framework of the TDF approach. In the model, the artificial coupling between the effective thermal diffusivity and heat capacity ratio as well as the artificial coupling between the effective mass diffusivity and porosity has been avoided, which is very useful in practical applications. Moreover, by adding source terms into the MRT-LB equations of the temperature and concentration fields, the macroscopic temperature and concentration equations can be correctly recovered. Numerical simulations of several benchmark problems are performed to demonstrate the accuracy and reliability of the present model. It is found that the present numerical results agree well with the well-documented data in the literature. In the present study, the Dufour and Soret effects have been neglected. For double-diffusive convection in porous media where the thermal and solutal buoyancies coexist, the Soret and Dufor effects may play important roles on the heat and mass transfer processes. The extension of the present model for simulating double-diffusive convection in porous media with Dufour and Soret effects is under way.

**Acknowledgments**

This work was financially supported by the Key Project of National Natural Science Foundation of China (Grant No. 51436007) and the Major Program of the National Natural Science Foundation of China (Grant No. 51590901).